\documentclass[letterpaper]{article} 
\usepackage{aaai24} 
\usepackage{times}  
\usepackage{helvet}  
\usepackage{courier}  
\usepackage[hyphens]{url}  
\usepackage{graphicx} 
\usepackage{natbib}  
\usepackage{caption} 
\usepackage{algorithm}
\usepackage{algorithmic}
\newenvironment{mechanism}[1][htb]{%
    \floatname{algorithm}{Mechanism}
   \begin{algorithm}[#1]%
  }{\end{algorithm}}

\usepackage{newfloat}
\usepackage{listings}
\DeclareCaptionStyle{ruled}{labelfont=normalfont,labelsep=colon,strut=off} 
\floatstyle{ruled}
\newfloat{listing}{tb}{lst}{}
\floatname{listing}{Listing}
\nocopyright

\usepackage{amsmath}
\usepackage{amsthm}
\usepackage{nicefrac}
\DeclareMathOperator*{\argmax}{arg\,max}

\newcommand{\ceil}[1]{\ensuremath{\left \lceil #1 \right \rceil}}
\newcommand{\floor}[1]{\ensuremath{\left \lfloor #1 \right \rfloor}}
\newtheorem{definition}{Definition}
\newtheorem{theorem}{Theorem}
\newtheorem{proposition}{Proposition}
\newtheorem{lemma}{Lemma}

\title{Social Mechanism Design: Making Maximally Acceptable Decisions}
\author{
    Ben Abramowitz,
    Nicholas Mattei
}
\affiliations{
    Tulane University\\

    babramow@tulane.edu, 
    nsmattei@tulane.edu
}

\begin{document}

\maketitle

\begin{abstract}
Agents care not only about the outcomes of collective decisions but also about \emph{how} decisions are made. In many cases, both the outcome and the procedure affect whether agents see a decision as legitimate, justifiable, or acceptable.
We propose a novel model for collective decisions that takes into account both the preferences of the agents and their higher order concerns about the process of preference aggregation.
To this end we (1) propose natural, plausible preference structures and establish key properties thereof, (2) develop mechanisms for aggregating these preferences to maximize the acceptability of decisions, and (3) characterize the performance of our acceptance-maximizing mechanisms.
We apply our general approach to the specific setting of dichotomous choice, and compare the worst-case rates of acceptance achievable among populations of agents of different types.
We also show in the special case of rule selection, i.e., amendment procedures, the method proposed by \citet{abramowitz2021beginning} achieves universal acceptance with certain agent types.
\end{abstract}

\section{Introduction}
In canonical mechanism design problems, a mechanism designer chooses an objective, makes assumptions about the preferences and behaviors of agents, and designs a method for aggregating the information received from agents into a ``good" solution \cite{maskin2008mechanism}. The quality of the outcome is measured by whatever metric the mechanism designer chooses. In this view, a good mechanism is one that produces good outcomes.

The economic and philosophical literature recognizes that this purely consequentialist view is often inappropriate. It is not irrational to find certain markets repugnant~\cite{roth2007repugnance}, hold that there are things money shouldn't buy~\cite{sandel2000money}, desire to be treated fairly~\cite{kahneman1986fairness}, or want that the exercise of power and authority be legitimate~\cite{sep-legitimacy}. In other words, quality of a mechanism is not solely a function of the quality of the outcomes it produces. This is borne out empirically, as voters often care whether their voting system achieves proportionality between parties~\cite{plescia2020people}.

In some settings, a choice of mechanism can be made separately from what outcomes it will produce in any single situation. For example, a group might form a constitution that specifies a voting rule to be used for all group decisions, irrespective of the decisions being made. Our focus here is on scenarios where the choice of mechanism or rule is tied up with the choice of outcome it produces in a particular instance.  

Our motivating observation is that the agents involved in collective decisions might care about all the same aspects that any mechanism designer or Social Choice theorist might. In many circumstances, one should assume that the agents involved are just as sophisticated, intelligent, knowledgeable, and conscientious as any mechanism designer (and generally more so). We must incorporate agent preferences over mechanisms to produce good mechanisms and make good decisions.

\paragraph{Contributions}
We propose and characterize several plausible, compact structures characterizing agent preferences over decisions.
These structures model agent preferences over decisions as consisting of (1) preferences over rules and outcomes, and (2) treatment of counterfactuals.
Based on the preference structures built on top of these properties we develop algorithms to aggregate such preferences into acceptance-maximizing decisions.
We provide a generic algorithm for computing acceptance maximizing decisions for all agent types, followed by more efficient, specialized algorithms when agents are all of one type. Subsequently, we apply our general framework to asymmetric dichotomous choice (ADC) problems, in which a proposal to be voted upon is put up against the status quo, and analyze the worst-case acceptance rate for each of our algorithms. We then extend our analysis to the procedures for rule updates or \emph{amendments} proposed by~\citet{abramowitz2021beginning}, where each amendment is an asymmetric dichotomous choice about whether to change the voting rule.

\section{Model}\label{sec:model}
 A set $N$ of agents with $|N| = n$ must make a collective decision. 
 A decision consists of applying some decision-making rule $R$ to select a single outcome $y$ based on the profile $V$ of information provided by the agents.
 Each agent $i \in N$ provides some information $v_i$, and these inputs collectively constitute a \emph{profile} $V = (v_1, \ldots, v_n)$.

 A decision consists of applying a function or \emph{rule} $R$ to a profile $V$, which yields an single outcome: $R(V) = y$.
 We denote a \emph{decision} by the tuple $(V, R, y)$ where $R(V) = y$. 
 \footnote{The notation $(V, R, y)$ is redundant since $R$ and $V$ imply outcome $R(V)$, but the verbose notation simplifies our presentation.}
 Let $\mathcal{V}$ be a set of profiles and $\bar{\mathcal{Y}}$ be a set of possible decision outcomes. We define $\bar{\mathcal{R}}$ to be the set of all functions $R$ such that $R(V) \in \bar{\mathcal{Y}}$ for all $V \in \mathcal{V}$.
 We can now define the set of possible decisions to be $\mathcal{D} = \{(V, R, y) : V \in \mathcal{V}, R \in \bar{\mathcal{R}}, y \in \bar{\mathcal{Y}}, \text{ and } R(V)=y \}$.

 Agents have approval preferences over decisions, i.e., every agent either accepts or does not accept a given decision. We denote by $\mathcal{D}_i \subseteq \mathcal{D}$ the subset of possible decisions an agent will accept and refer to this as the agent's \emph{satisfying set}. We denote the agents' satisfying sets collectively by $\mathcal{D}_N$. Note that $\mathcal{D}_N$ constitutes additional information about preferences over decisions not contained in $V$, so the total information provided by the voters is $(V, \mathcal{D}_N)$. 

 \smallskip
 \noindent
 \textbf{Feasibility.}
 In a single problem instance a single profile $V \in \mathcal{V}$ is observed and a single decision must be made, but not all of the decisions in $\mathcal{D}$ may be feasible. There may be  constraints, e.g. what rules can be implemented in a timely manner or what outcomes there is sufficient budget to afford.
 Feasibility constraints on rules and outcomes restrict what decisions can be made, but these are not constraints on profiles or satisfying sets. The feasibility constraints may not even be known to the agents. For example, in participatory budgeting agents might not have full knowledge of the budget constraints.

 For a given instance, within the set of possible outcomes, there is a subset of feasible outcomes $\mathcal{Y} \subseteq \bar{\mathcal{Y}}$, and within the set of all possible rules, there is a subset of feasible rules $\mathcal{R} \subseteq \bar{\mathcal{R}}$. A decision $(V, R, y)$ is a \emph{feasible decision} if and only if $R \in \mathcal{R}$ and $y \in \mathcal{Y}$, and we require that our agents must always make a feasible decision.

 In some instances, a feasible rule may not yield a feasible outcome when applied to the profile. It is possible that $R(V) \not \in \mathcal{Y}$ for some $R \in \mathcal{R}$. Likewise, a feasible outcome may not be selected as the winning outcome by any feasible rule for some profiles. For the decision $(V, R, y)$ to be feasible, both the rule and outcome are required to be independently feasible; $R \in \mathcal{R}$ and $y \in \mathcal{Y}$.

 \smallskip
 \noindent
 \textbf{Acceptance Maximization.}
 We denote a single instance by $I = (V, \mathcal{R}, \mathcal{Y}, \mathcal{D}_N)$, in which the relevant information about $\mathcal{V}, \bar{\mathcal{R}}$, and $\bar{\mathcal{Y}}$ is implicit in $\mathcal{R}$, $\mathcal{Y}$, and $\mathcal{D}_N$. For any collection of instances $\mathcal{I}$, we would like to specify a meta-rule or \emph{mechanism} $M$ that returns an acceptance-maximizing feasible decision for all instances in that collection.
 
 We make a single assumption that precludes indeterminacy: For every problem instance $I = (V, \mathcal{R}, \mathcal{Y}, \mathcal{D}_N)$, $\exists R \in \mathcal{R}$ such that $R(V) \in \mathcal{Y}$. That is, for all instances we consider, at least one feasible decision exists.
 Our goal is to select a feasible decision accepted by the greatest number of agents. If there are multiple acceptance-maximizing decisions, one can break ties arbitrarily (e.g., randomly).

 Let $\mathcal{F}^I \subseteq \mathcal{D}$ be the set of all feasible decisions in instance $I$.
 For each instance $I$, the acceptance-maximizing mechanism $M$ yields a decision $M(I) \in \mathcal{F}^I$.
 $$M(I) = \argmax\limits_{(V, R, y) \in \mathcal{F}^I} |\{i \in N: (V, R, y) \in \mathcal{D}_i\}|$$

 Given a collection of instances $\mathcal{I}$, we are primarily concerned with the best worst-case acceptance rate achievable by an acceptance-maximizing mechanism $M$.
 $$\alpha_{\mathcal{I}} := \min\limits_{I \in \mathcal{I}} \frac{|\{i \in N: M(I) \in \mathcal{D}_i\}|}{n}$$

\smallskip
 \noindent
 \textbf{Elicitation.}
 For the moment, while introducing our general model, we are not concerned with the nature of the outcomes or what information the profile contains, e.g. whether it consists of categorical, ordinal, or cardinal information. In Sections \ref{sec:adc} and \ref{sec:amendment} we will specify explicit domains for the sets $\mathcal{V}$, $\mathcal{R}$, and $\mathcal{Y}$. While $V$ can be any sort of information in our general model, we will focus on social choice settings where $v_i \in \{0,1\}$ reflects agent $i$'s preference between two outcomes.

 In general, the set of possible decisions $\mathcal{D}$ can be intractably large or even infinite depending on the cardinalities of $\mathcal{V}$ and $\bar{\mathcal{Y}}$. In such cases, it may be unrealistic to assume that agents simply provide a list of all decisions they accept. Rather, we investigate satisfying sets $\mathcal{D}_i$ that are structured and amenable to compact representation. Our work centers on the identification of realistic structures for satisfying sets and the features that give them such structure.

 We separate $v_i$ from $\mathcal{D}_i$ but this information will often be expressed together. Often we might expect $v_i$ and $\mathcal{D}_i$ to relate to each other; i.e. I will accept the outcome if the candidate I voted for wins. However, we do not assume that agents necessarily know the profile $V$ when they report $\mathcal{D}_i$, e.g. a voter does not need to know the preferences of all other voters. Thus, in any problem instance, $\mathcal{D}_i$ will typically contain information about many profiles other than the single profile that gets observed, and this information will typically be ignored by our acceptance-maximizing mechanism.

 \subsection{Agent Types}
We associate an agent's type with certain structure in their satisfying set. The structures of satisfying sets allow for (relatively) compact representation and have intuitive meanings. To illustrate the different agent types we use a toy example of a group of friends are deciding where to go for dinner between three restaurants $\mathcal{Y} = \{A, B, C\}$. 
Acceptance of the decision means willingness to attend the dinner. We can imagine different ways in which agents might communicate which decisions $\mathcal{D}_i$ they would accept. 
Each time we define an agent type we will give an example of what an agent of that type might report when deciding on a restaurant.

A consequentialist only cares about the outcome, e.g. ``I will go to A or B, but not C" while an absolute proceduralist only cares about the rule, e.g. ``I will only go if I get to choose the restaurant this time."

\begin{definition}[Consequentialism]
Agent $i$ is consequentialist if there exists a set of outcomes $Y_i \subseteq \bar{\mathcal{Y}}$ such that for all decisions $(V, R, y) \in \mathcal{D}$, $(V, R, y) \in \mathcal{D}_i$ if and only if $y \in Y_i$.
\end{definition}

\begin{definition}[Absolute Proceduralism]
Agent $i$ is absolute proceduralist if $\exists R_i \subseteq \bar{\mathcal{R}}$ such that $\forall (V, R, y) \in \mathcal{D}$, $(V, R, y) \in \mathcal{D}_i$ if and only if $R \in R_i$
\end{definition}

There are two basic ways agents might combine their concerns about the outcome and the rule -- conjunction and disjunction. An absolute conjunctivist might say, ``We should use Plurality voting to decide, but I will never go to restaurant B" while an absolute disjunctivist might say, ``I will go to whichever restaurant wins by Plurality voting, but I am always happy going to restaurant A regardless of how we decide."

\begin{definition}[Absolute Conjunctivism]
Agent $i$ is an absolute disjunctivist if $\exists R_i \subseteq \bar{\mathcal{R}}$ and $Y_i \subseteq \bar{\mathcal{Y}}$ such that $(V, R, y) \in \mathcal{D}_i$ if and only if $R \in R_i$ \textbf{and} $y \in Y_i$
\end{definition}

\begin{definition}[Absolute Disjunctivism]
Agent $i$ is an absolute disjunctivist if $\exists R_i \subseteq \bar{\mathcal{R}}$ and $Y_i \subseteq \bar{\mathcal{Y}}$ such that $(V, R, y) \in \mathcal{D}_i$ if and only if $R \in R_i$ \textbf{or} $y \in Y_i$
\end{definition}

Consequentialism and absolute proceduralism are sub-types of absolute disjunctivism in which $R_i = \emptyset$ or $Y_i = \emptyset$, respectively.
If two agents have the same sets $R_i, Y_i$ for a given profile, but one is an absolute disjunctivist and the other is an absolute conjunctivist, the disjunctivist is easier to satisfy because they will accept more decisions.

In any instance, a decision $(V, R, y)$ is made in which exactly one rule $R$ is implemented. However, there may be other rules $R'$ such that $R'(V) = y$. When only $V$ and $y$ are observable, one cannot necessarily infer what rule was implemented. Reasoning about which rule was implemented requires counterfactual reasoning about what the outcome would have been had the profile been different. 
If two agents cannot observe the rule they might disagree about what decision was made, even though they have common knowledge of the profile and outcome. If two (absolutist) agents $i,j \in N$ have identical satisfying sets $\mathcal{D}_i = \mathcal{D}_j$ but disagree about whether the decision made was $(V, R, y)$ or $(V, R', y)$ they may not agree on whether to accept the decision. 

We now define corresponding agent types for whom the counterfactuals do not influence acceptance. We refer to an agent as \emph{implementation-indifferent} if we can determine whether they accept a decision $(V, \_\_, y)$ without observing what rule was implemented. Note carefully that implementation-indifference does not mean an agent is a consequentialist, and they may still express their preferences over decisions in terms of rules.

\begin{definition}[Implementation-Indifference]
Agent $i$ is implementation-indifferent (II) if for all decisions $(V, R, y), (V, R', y) \in \mathcal{D}$, we have $(V, R, y) \in \mathcal{D}_i \Rightarrow (V, R', y) \in \mathcal{D}_i$.
\end{definition}

Preferences over decisions for selecting a restaurant might sound like the following. AN II-Proceduralist might say, ``I will go wherever the most people want to go", and II-Conjunctivist might say, ``I will go anywhere except for restaurant A, as long as it's where the most people want to go", and an II-Disjunctivist could say ``I will go anywhere except for restaurant A, unless everyone else wants to go to A, then I'll go too."

\begin{definition}[II-Conjunctivism]
Agent $i$ is an II-conjunctivist if $\exists R_i \subseteq \bar{\mathcal{R}}$ and $Y_i \subseteq \bar{\mathcal{Y}}$ such that $\forall (V, R, y) \in \mathcal{D}$, we have $(V, R, y) \in \mathcal{D}_i$ if and only if $y \in Y_i$ \textbf{and} $\exists R^* \in R_i$ such that $R^*(V) = y$.
\end{definition}

\begin{definition}[II-Disjunctivism]
Agent $i$ is an II-disjunctivist if $\exists R_i \subseteq \bar{\mathcal{R}}$ and $Y_i \subseteq \bar{\mathcal{Y}}$ such that $\forall (V, R, y) \in \mathcal{D}$, we have $(V, R, y) \in \mathcal{D}_i$ if and only if $y \in Y_i$ \textbf{or} $\exists R^* \in R_i$ such that $R^*(V) = y$.
\end{definition}

Consequentialist agents are always implementation-indifferent, and an II-Proceduralist is an II-Disjunctivist with $Y_i = \emptyset$.
All satisfying sets we have defined can be characterized by a tuple $\mathcal{D}_i = (R_i, Y_i, \Psi_i, \Phi_i)$ where $\Psi_i$ labels a satisfying set as conjunctivist (1) or disjunctivist (0), and $\Phi_i$ indicates implementation-indifference (1) or non-II (0).\footnote{A satisfying set should properly be defined with respect to a collection of instances. We drop this dependence in our notation.} For brevity, we assume going forward that all $R_i$ and $Y_i$ are finite sets.

\section{Generic Acceptance Maximization}
We now demonstrate how to make acceptance-maximizing decisions for all agent types, starting with absolute disjunctivists.

\begin{mechanism}[h]
\caption{\small Max Acceptance for Absolute Disjunctivists}
\begin{algorithmic} \label{alg:abs_disj}
\REQUIRE $(V, \mathcal{R}, \mathcal{Y}, \mathcal{D}_N)$ 
\FOR{$a \in \mathcal{Y}$}
    \STATE $N^a \gets \{i \in N : a \in Y_i\}$
    \STATE $R^a \gets \argmax\limits_{R \in \mathcal{R} : R(V) = a} |\{i \in N \backslash N^a : R \in R_i\}|$
\ENDFOR
\STATE $y = \argmax\limits_{a \in \mathcal{Y}: R^a \in \mathcal{R}}  |N^a \cup \{i \not \in N^a : R^a \in R_i\}|$
\RETURN $(V, R^y, y)$
\end{algorithmic}
\end{mechanism}

\begin{theorem}\label{thm:max_abs_disj}
    Mechanism~\ref{alg:abs_disj} maximizes acceptance among feasible decisions for absolute disjunctivists.
\end{theorem}

\begin{proof}[Proof of Theorem \ref{thm:max_abs_disj}]
    For each feasible outcome $a \in \mathcal{Y}$, Mechanism~\ref{alg:abs_disj} finds the set of agents $N^a$ for whom $a \in Y_i$ and finds the feasible rule $R^a$ that yields this outcome on the given profile for which the maximum number of agents not in $N^a$ have $R^a \in R_i$. For each outcome $a$ it therefore maximizes the number of agents who are either in $N^a$ or have $R^a \in R_i$. The rule and outcome $R^y$ and $y$ are then selected to be the ones that maximize the number of agents who accept the decision based on one of those two reasons. $R^a$ does not exist if no feasible rule selects $a$ on the profile.
\end{proof}

In a single instance, once profile $V$ is observed, the relevant decisions in $\mathcal{D}_i$ are only those containing $V$. Using this observation, we can build a general mechanism for all agent types on top of Mechanism~\ref{alg:abs_disj} by finding for every satisfying set $\mathcal{D}_i$ a substitute absolute disjunctivist satisfying set $\tilde{\mathcal{D}}_i$ that accepts the same set of decisions among those that are feasible and contain the profile $V$.

\begin{lemma}\label{lem:equiv_abs_disj}
    $\forall \mathcal{D}_i \subseteq \mathcal{D}$, $\forall V^* \in \mathcal{V}$, $\exists \tilde{R}_i \subseteq \mathcal{R}, \tilde{Y}_i \subseteq \mathcal{Y}$ such that $(V^*, R, y) \in \mathcal{D}_i$ if and only if $R \in \tilde{R}_i$ or $y \in \tilde{Y}_i$.
\end{lemma}

\begin{proof}[Proof of Lemma \ref{lem:equiv_abs_disj}]
    For II-disjunctivists, for every rule $R$ in $R_i$ add $y=R(V)$ to form $\tilde{Y}_i$ and then set $\tilde{R}_i = \emptyset$, and $\tilde{D}_i = (\tilde{R}_i, \tilde{Y}_i, 0, 0)$. Now $(V, R, y) \in \tilde{D}_i$ if and only if $y \in \tilde{Y}_i$ which requires $y \in Y_i$ or $\exists R \in R_i$ such that $R(V) = y$.
    For II-conjunctivists, we construct $\tilde{Y}_i$ by removing each outcome $y$ from $Y_i$ if $\not\exists R \in R_i$ such that $R(V)=y$, set $R_i = \emptyset$, and $\tilde{D}_i = (\tilde{R}_i, \tilde{Y}_i, 0, 0)$. Now $(V, R, y) \in \tilde{D}_i$ if and only if $y \in \tilde{Y}_i$ which requires $y \in Y_i$ and $\exists R \in R_i$ such that $R(V) = y$.
    Lastly, for absolute conjunctivists, create $\tilde{R}_i$ by removing from $R_i$ any rule $R$ such that $R(V) \not\in Y_i$, set $\tilde{Y}_i = \emptyset$, and $\tilde{D}_i = (\tilde{R}_i, \tilde{Y}_i, 0, 0)$.  Now $(V, R, y) \in \tilde{D}_i$ if and only if $R \in \tilde{R}_i$ which requires that $R \in R_i$ and $R(V) \in Y_i$.
\end{proof}

From Lemma \ref{lem:equiv_abs_disj} and Theorem \ref{thm:max_abs_disj} it follows that we can maximize acceptance among any mix of agents of the types defined in Section~\ref{sec:model}. For any profile $V$ our mechanism $M$ can first compute substitute absolute disjunctivist satisfying sets for all agents that accept the same subset of feasible decisions for profile $V$, and then apply Mechanism \ref{alg:abs_disj}.

\begin{mechanism}[h!]
\caption{\small Max Acceptance with All Agent Types}
\label{alg:all}
\begin{algorithmic} 
\REQUIRE $(V, \mathcal{R}, \mathcal{Y}, \mathcal{D}_N)$
\FOR{$i \in N$}
    \STATE $\tilde{Y}_i \gets \emptyset$
    \STATE $\tilde{R}_i \gets \emptyset$
    \IF{$i$ is an II-disjunctivist}
            \STATE $\tilde{Y}_i \gets Y_i$
            \FOR{$R \in R_i$}
                \STATE $\tilde{Y}_i \gets \tilde{Y}_i \cup R(V)$
            \ENDFOR
    \ELSIF{$i$ is II-conjunctivist}
        \FOR{$R \in R_i$}
            \IF{$R(V) \in Y_i$}
                \STATE $\tilde{Y}_i \gets \tilde{Y}_i \cup R(V)$
            \ENDIF
        \ENDFOR
    \ELSIF{$i$ is an absolute conjunctivist}
        \FOR{$R \in R_i$}
            \IF{$R(V) \in Y_i$}
                \STATE $\tilde{R}_i \gets \tilde{R}_i \cup R$
            \ENDIF
        \ENDFOR
    \ENDIF
    \STATE $\tilde{\mathcal{D}}_i \gets (\tilde{R}_i, \tilde{Y}_i, 0, 0)$
\ENDFOR
\RETURN Mechanism~\ref{alg:abs_disj}$(V, \mathcal{R}, \mathcal{Y}, \tilde{\mathcal{D}}_N)$
\end{algorithmic}
\end{mechanism}

The process of creating substitute satisfying sets in Mechanism~\ref{alg:all} is clearly inefficient when it requires computing the outcomes for a large number of feasible rules. When all agents are of the same type and non-II, we provide simpler, more direct mechanisms for computing acceptance-maximizing decisions (relatively) efficiently in the appendix. 
Once we specify concrete domains for $V$, $\bar{\mathcal{R}}$, and $\bar{\mathcal{Y}}$ in the following sections, we will construct efficient algorithms particular to those domains.

Intuitively, disjunctivists accept more decisions than consequentialists and proceduralists who each accept more decisions than conjunctivists, and II agents will tend to accept more decisions than non-II agents. Despite how generic our approach has been so far, we can start to establish lower bounds on the achievable acceptance rate. 

\begin{proposition}\label{prop:II_all}
    If all agents are implementation-indifferent and for each agent, for every profile, there exists at least one feasible decision they accept, then $\alpha_\mathcal{I} \geq \frac{1}{|\mathcal{Y}|}$.
\end{proposition}

\begin{proof}
    When all agents are II, we can determine if they accept a decision $(V, \_\_, y)$ regardless of the rule. Given $V$, the number of possible tuples $(V,y)$ is $|\mathcal{Y}|$. With $n$ agents and $|\mathcal{Y}|$ feasible outcomes, for at least one outcome there must be at least $\frac{n}{|\mathcal{Y}|}$ agents who will accept any decision with that outcome by the pigeonhole principle. 
\end{proof}

\begin{table*}[h!]
\begin{center}
\begin{tabular}{||c c c||} 
 \hline
 Agent Type & Assumptions $\forall i \in N$ & $\alpha_\mathcal{I}$ \\ [0.5ex] 
 \hline\hline
 Any & None & 0 \\ 
 \hline
 Absolute Conjunctivists & $v_i \in Y_i$ and $|R_i\cap \mathcal{R}| \geq 1$ & 0 \\
 \hline
 Absolute Conjunctivists & $|Y_i \cap \mathcal{Y}| \geq 1$ and $\exists R \in R_i \cap \mathcal{R} : R(V) \in Y_i \cap \mathcal{Y}$ & $\nicefrac{2}{n}$ \\
 \hline
 Absolute Disjunctivists & $|R_i \cap \mathcal{R}| \geq 1$ & $\nicefrac{2}{n}$ \\
 \hline
 Absolute Disjunctivists & $|R_i\cap \mathcal{R}| \geq k$ & $\frac{1}{n} \cdot \ceil{\frac{nk}{\floor{\frac{n+1}{2}}}}$\\
 \hline
  Absolute Disjunctivists & $|Y_i \cap \mathcal{Y}| \geq 1$ & $\nicefrac{1}{2}$ \\ 
 \hline
 II-Conjunctivists & $|Y_i \cap \mathcal{Y}| \geq 1$ and $\exists R \in R_i : R(V) \in Y_i \cap \mathcal{Y}$ & $\nicefrac{1}{2}$ \\
 \hline
 II-Disjunctivists & $\exists R \in R_i : R(V) \in \mathcal{Y}$ & $\nicefrac{1}{2}$ \\
 \hline
 II-Disjunctivists & $|Y_i \cap \mathcal{Y}| \geq 1$ & $\nicefrac{1}{2}$ \\
 \hline
 II-Disjunctivists & $|Y_i \cap \mathcal{Y}| = 1$ and $\exists R \in R_i : R(V) \in \mathcal{Y} \backslash Y_i$ & 1 \\ [1ex] 
 \hline
\end{tabular}
\caption{Worst-case acceptance rates for ADC problems with all agents of one type, given assumptions about $V$ and $\mathcal{D}_N$.}
\label{table:ADC}
\end{center}
\end{table*}

\section{Asymmetric Dichotomous Choice}\label{sec:adc}
We move now from the generic model to the concrete, reality-aware setting of choosing between two alternatives~\cite{shapiro2017reality,shahaf2018sybil}.
For instances of asymmetric dichotomous choice (ADC) problems, the feasible outcomes are $\mathcal{Y} = \{r,p\}$ where $r$ is the status quo and $p$ is a competing proposal.
Each agent must express a vote $v_i \in \{r,p\}$, so the set of possible profiles is $\mathcal{V} = \{r,p\}^n$ with $n$ agents.
The set of feasible rules $\mathcal{R}$ we allow is the set of supermajority rules. We define a supermajority rule, denoted $R^\delta$ where $\frac{1}{2} \leq \delta < 1$, as a rule that chooses $p$ as the outcome if and only if strictly greater than $\delta n$ agents vote for it; otherwise selecting $r$.
Here, $R^\frac{1}{2}$ is the majority rule which breaks ties in favor of the status quo, and $R^{1 - \frac{1}{n}}$ is unanimity rule.
There are exactly $|\mathcal{R}| = \lfloor \frac{n+1}{2} \rfloor$ distinct supermajority rules with $n$ agents.
To simplify notation we refer interchangeably to a rule $R^\delta$ and its threshold $\delta$.

We now identify efficient mechanisms for acceptance maximization in collections $\mathcal{I}$ of ADC problem instances with homogeneous agents of each type, and give worst-case bounds on the acceptance rate under various assumptions on the structure of $V$ and $\mathcal{D}_N$. For example, it is often reasonable to assume a relationship between $v_i$ and $\mathcal{D}_i$ for each agent, e.g. $v_i \in Y_i$. Below we provide mechanisms for consequentialists, absolute disjunctivists, and II-disjunctivists for illustration and include a proof of acceptance maximization for consequentialists. The remaining mechanisms for the other agent types as well as proofs that they maximize acceptance can be found in the appendix.
The acceptance rates under various assumptions are given in Table~\ref{table:ADC}, and concretely demonstrate the intuition that (1) acceptance tends to be considerably higher among II agents (vs. non-II agents) and disjunctivists (vs. conjunctivists). The assumptions in Table~\ref{table:ADC} are minimal, e.g. that for each agent there is at least one feasible decision they accept. The proofs of each bound in Table~\ref{table:ADC} can be found in the appendix.

\begin{mechanism}[h!]
\caption{\small Max Acceptance for ADC with Consequentialists}
\begin{algorithmic} \label{alg:ADC_Abs_Conseq}
\REQUIRE $(V, \mathcal{R}, \mathcal{Y}, \mathcal{D}_N)$ 

\STATE $V_r \gets |\{i \in N : v_i = r\}|$
\STATE $V_p \gets |\{i \in N : v_i = p\}|$
\STATE $N_r \gets |\{i \in N : r \in Y_i\}|$
\STATE $N_p \gets |\{i \in N : p \in Y_i\}|$

\IF{$V_p = |N|$}
    \RETURN $(V, R^{1 - \frac{1}{n}}, p)$
\ELSIF{$N_r \geq N_p$ \OR $V_r \geq V_p$}
    \RETURN $(V, R^{1 - \frac{1}{n}}, r)$
\ELSE
    \RETURN $(V, R^\frac{1}{2}, p)$
\ENDIF

\end{algorithmic}
\end{mechanism}

\begin{proposition}\label{prop_adc_abs_conseq_max}
    Mechanism~\ref{alg:ADC_Abs_Conseq} maximizes acceptance for all ADC instances with consequentialist agents. 
\end{proposition}

\begin{proof}
    If status quo $r$ receives at least as many votes as proposal $p$, no supermajority rule will select $p$ so the only feasible decisions have outcome $r$, and the decision can be made by any supermajority rule. Oppositely, if proposal $p$ receives more votes than $r$ and $N_p > N_r$, we maximize acceptance with a feasible decision that selects $p$. Majority rule selects $p$, so the decision $(V, R^\frac{1}{2}, p)$ is feasible and maximizes acceptance.
    If $N_r \geq N_p$, a feasible decision that maintains the status quo as the outcome will satisfy at least as many agents as a feasible decision with the proposal as the outcome. Unanimity rule will always select the status quo unless the agents unanimously voted for the proposal.
    Lastly, if $N_r \geq N_p$ but the agents unanimously voted for the proposal, no feasible decision selects $r$. The outcome must be $p$, and the decision can be made by any supermajority rule.
\end{proof}

Consequentialists ignore the rule but Mechanism~\ref{alg:ADC_Abs_Conseq} must ensure the rule is feasible. By contrast, with absolute disjunctivists, the choice of rule matters for the outcome it selects, ensuring feasibility, and its presence in satisfying sets.
Recall agents may have infeasible decisions in their satisfying sets, and may not even be aware of the feasibility constraints. In Mechanism \ref{alg:ADC_Abs_Disj}, for non-II agents, only the feasible rules in $R_i$ are relevant for maximizing acceptance, but this is not the case for II agents, as reflected in Mechanism \ref{alg:II_binary}.

\begin{mechanism}[h!]
\caption{\small Max Acceptance for ADC with Abs. Disjunctivists}
\begin{algorithmic} \label{alg:ADC_Abs_Disj}
\REQUIRE $(V, \mathcal{R}, \mathcal{Y}, \mathcal{D}_N)$ 

\STATE $\delta_r = \frac{|\{i \in N : v_i = p\}|}{|N|}$
\STATE $N_r \gets \{i \in N : r \in Y_i\}$
\STATE $N_p \gets \{i \in N : p \in Y_i\}$
\STATE $R^*_r \gets \argmax\limits_{R^\delta \in \mathcal{R} : \delta \geq \delta_r} |\{i \in N \backslash N_r: R^\delta \in R_i\}|$
\STATE $R^*_p \gets \argmax\limits_{R^\delta \in \mathcal{R} : \delta < \delta_r} |\{i \in N \backslash N_p: R^\delta \in R_i\}|$

\STATE $N_r \gets N_r \cup \{i \in N : R^*_r \in R_i\}$
\STATE $N_p \gets N_p \cup \{i \in N : R^*_p \in R_i\}$

\IF{$|N_r| \geq |N_p|$}
    \RETURN $(V, R^*_r, r)$
\ELSE
    \RETURN $(V, R^*_p, p)$
\ENDIF

\end{algorithmic}
\end{mechanism}

\begin{mechanism}[h!]
\caption{\small Max Acceptance for ADC with II-Disjunctivists}
\begin{algorithmic} \label{alg:II_binary}
\REQUIRE $(V, \mathcal{R}, \mathcal{Y}, \mathcal{D}_N)$ 
\STATE $\delta_r = \frac{|\{i \in N : v_i = p\}|}{|N|}$
\STATE $\delta_p = \frac{|\{i \in N : v_i = p\}|-1}{|N|}$
\STATE $N_r \gets \{i \in N : r \in Y_i \texttt{ \OR} \max(R_i) \geq \delta_r \}$
\STATE $N_p \gets \{i \in N : p \in Y_i \texttt{ \OR} \min(R_i) \leq \delta_p \}$
\IF{$|N_r| \geq |N_p|$}
    \RETURN $(V, R^{\delta_r}, r)$
\ELSE
    \RETURN $(V, R^{\delta_p}, p)$
\ENDIF
\end{algorithmic}
\end{mechanism}

\section{Supermajority Rule Selection (Amendment)}\label{sec:amendment}
 An amendment is a special case of an asymmetric dichotomous choice where the potential outcomes are also rules~\cite{abramowitz2021beginning}. 
 Let $\mathcal{R}$ be the set of supermajority rules with $n$ agents as previously defined.
 As before, we refer interchangeably to a rule $R^\delta$ and its threshold $\delta$.
 For each amendment problem, there is a status quo, or supermajority rule $r \in \mathcal{R}$ currently in use.
 The agents are to decide whether or not to change the rule to be $p \in \mathcal{R}$, so $\mathcal{Y} = \{r,p\}$.
 Each agent casts a vote $v_i \in \{r,p\}$ forming collective profile $V_{rp}$.
 Since the votes $v_i$ are rules, we assume a correspondence between the profile and the satisfying sets of the agents.
 Each agent $i \in N$ has a strict preference ordering $\succ_i$ over $\mathcal{R}$. We denote the most preferred rule at the top of $i$'s ordering by $\delta_i$, and assume that agent preferences are single-peaked over the numerical order of $\mathcal{R}$ with the peak at $\delta_i$. In other words, for any two supermajority rules $\delta_1, \delta_2 \in \mathcal{R}$, if $\delta_1 < \delta_2 < \delta_i$ then $\delta_2 \succ_i \delta_1$ and if $\delta_i < \delta_1 < \delta_2$ then $\delta_1 \succ_i \delta_2$.
 Between any two rules, agents vote for the one they prefer in $V_{rp}$. With status quo $r$ and proposed rule $p$, if $r \succ_i p$ then we assume $v_i = r$, and similarly, if $p \succ_i r$ then $v_i = p$.
 Following~\citet{abramowitz2020amend}, our focus is on II-disjunctivists voting on amendments. We make two assumptions about how the profile corresponds to their satisfying sets: for all $i \in N$, $v_i \in Y_i$ and $\delta_i \in R_i$.
 In other words, an agent accepts the outcome of an amendment whenever their preferred outcome wins and whenever their ideal rule would have selected the same outcome on the profile $V_{rp}$.

\smallskip
\noindent
\textbf{Universal Acceptance.}
One would hope that if the decision made for choosing a rule reaches universal acceptance $(\alpha_I=1)$ then all future decisions using the chosen rule will be unanimously accepted on procedural grounds~\cite{dietrich2005reach}.
Since we have assumed our agents are II-disjunctivists, we can use Mechanism~\ref{alg:II_binary} to compute an asymmetric binary decision that maximizes acceptance.
Unfortunately, for some instances there is no decision that achieves universal acceptance.
Given status quo $r$, we want to know for what proposal $p$ and amendment rule $R \in \mathcal{R}$ does the decision $(V_{rp}, R, R(V_{rp}))$ maximize acceptance.
Amazingly, given any status quo $r < 1 - \frac{1}{n}$, there exists at least one proposal $p \neq r$, and amendment rule $R$, such that $(V_{rp}, R, R(V_{rp}))$ is guaranteed to achieve universal acceptance for any profile $V_{rp}$ induced by agents' single-peaked preferences over supermajority rules.

\begin{lemma}\label{lemma:hrule_increment}
    The amendment decision $(V_{r,r+\frac{1}{n}}, R^r, R^r(V_{r,r+\frac{1}{n}}))$ is universally accepted by II-disjunctivists with single-peaked preferences over supermajority rules ordered on the real line such that $v_i \in \{r,p\}$, $v_i \in Y_i$, and $\delta_i \in R_i$ for all agents.
\end{lemma}

\begin{proof}
Let $p = r+\frac{1}{n}$. Suppose $R^r(V_{rp}) = r$. Then for all $i \in N$ such that $\delta_i \geq r$, $(V_{rp}, R^r, r) \in \mathcal{D}_i$ because $R^{\delta_i}$ yields the same outcome $r$. For all $i \in N$ such that $\delta_i < r$, $r \in Y_i$.
Now suppose $R^r(V_{rp}) = p$. Then for all $i \in N$ such that $\delta_i \leq r$, $(V_{rp}, R^r, p) \in \mathcal{D}_i$ because $R^{\delta_i}$ yields  the same outcome $p$. For all $i \in N$ such that $\delta_i > r$, $p \in Y_i$.
\end{proof}

Based on Lemma \ref{lemma:hrule_increment}, we can build an iterative algorithm, Mechanism~\ref{alg:hrule}, for considering sequences of proposals to change the supermajority rule in incrementally increasing order such that each decision along the way is universally accepted.
Mechanism~\ref{alg:hrule} does not consider all possible proposals. It only considers those greater than $r$. When $r = \frac{1}{2}$ all other supermajority rules have the potential to be proposed. There are additional benefits to ensuring the original status quo be $\frac{1}{2}$, or at least that it not be too large. 
The following special supermajority rule is labeled $h$ due to its resemblance to the $h$-index in bibliometrics:
$$h = \argmax\limits_{p \in \mathcal{R}} |\{i \in N: \delta_i \geq p\}| \geq np$$

\begin{mechanism}[h!]
\caption{\small Supermajority Rule Amendment}
\begin{algorithmic} \label{alg:hrule}
\REQUIRE $(V, \mathcal{R}, \mathcal{Y}, \mathcal{D}_N)$ 
\STATE $p \gets r+\frac{1}{n}$
\vspace{2mm}
\WHILE{$p < 1$}
    \IF{$R^r(V_{rp}) = p$}
        \STATE $r \gets p$
        \STATE $p \gets p + \frac{1}{n}$
    \ELSE
        \RETURN $(V, R^r, R^r(V))$
    \ENDIF
\ENDWHILE
\end{algorithmic}
\end{mechanism}

\begin{theorem}\label{thm:hrule_each_step}
    If $r \leq h$, then Mechanism~\ref{alg:hrule} returns $(V, R^h, h)$, otherwise it returns $(V, R^r, r)$.
\end{theorem}

\begin{proof}
Assume $r < p \leq h$ where $p = r + \frac{1}{n}$. If $R^r(V_{rp}) = r$, there are fewer than $pn$ with peaks $\delta_i \geq p$, but this violates the definition of $h$, since there must be at least $hn > pn$ agents with peaks $\delta_i \geq h \geq p$.
Thus, all amendments to increment the status quo will be successful if $r < h$.
If $p > r \geq h$, there are at most $rn$ agents with $\delta_i > r$ by the definition of $h$ and all other agents must prefer $r \succ_i p$, so no proposal $p > r$ can possibly succeed as an amendment.
\end{proof}

One of the constitutions proposed by \citet{abramowitz2021beginning}, is derived from axioms that uniquely imply the initial supermajority rule when founding the constitution should be $r = \frac{1}{2}$, and then Mechanism~\ref{alg:hrule} should be applied. Our analysis shows a different perspective; when agents are II-disjunctivists with votes and satisfying sets based on $\delta_i$ values, every amendment decision can be universally accepted under such a constitution.

Implementing Mechanism~\ref{alg:hrule} with its iterative, incremental changes to the status quo is tedious. Do the agents need to observe every update to the status quo?
Suppose we know all agents' ideal rules $\delta_i$, and can therefore infer a partial ordering consistent with their preference ordering over $\mathcal{R}$ based on single-peakedness. Let this collection of partial orderings be the profile $V$.
If we can compute $h$ directly from $V$, and $r < h$, we could directly implement the decision $(V, R, h)$ with the rule that always selects $h$. This can be achieved by implementing the amendment decision $(V_{rh}, R^r, h)$. We know that any further proposals to amend $h$ will fail, and the decision to maintain $h$ against any proposal will be universally accepted.
This more efficient algorithm corresponds to the second amendment procedure proposed by \citet{abramowitz2021beginning}, derived from a similar set of axioms to the first.
What happens to the acceptance rate for the amendment decision $(V_{rh}, R, h)$? Amazingly, $\alpha_{\mathcal{I}}=1$.

\begin{theorem}\label{thm:hrule_one_step}
    If the status quo is $r < h$, the amendment $(V_{rh}, R^r, h)$ is universally accepted by II-disjunctivists with $v_i \in Y_i$ and $\delta_i \in R_i$.
\end{theorem}

\begin{proof}
    Assume $r < h$.
    Any agent who prefers $h \succ_i r$ will accept the decision $(V_{rh}, R^r, h)$ because $h \in Y_i$.
    For all $i \in N$ such that $r \succ_i h$, we know that $\delta_i < h$ due to single-peakedness.
    We also know there must be at least $hn$ agents prefer $h \succ_i r$ by the definition of $h$.
    Thus, for any agent for whom $r \succ_i h$, they accept $R^r(V_{rh}) = h$ because $R^{\delta_i}$ would yield the same outcome.
\end{proof}


\section{Related Work}
Recently,~\citet{procaccia2019axioms} proposed that we should use axioms to ``help explain the mechanism’s outcomes to participants." The main reason given is that ``explanations make the solutions more appealing to users, and, therefore, make it more likely that users will \textbf{accept} them."
Automated procedures have been developed for providing agents with axiomatic justifications for decisions which consist of an explanation and a set of axioms that serve as a ``normative basis"~\cite{boixel2020automated}.
We would like to provide as many agents as possible with a personalized justification they accept, which may differ from the justification given to other agents.
After all, part of what makes collective decisions practical is that different people can accept the same decision for different reasons.
If we know what justifications each agent will accept, we can turn the problem on its head and compute decisions that are justifiable to the maximum number of agents.
The personalization of justifications raises important questions about what it means for multiple justifications to be compatible with one another. 
 Our model formalizes one aspect of how agents might view the relevant counterfactuals with our notion of \emph{implementation-indifference}, which is related to the difference between intra-profile and inter-profile axioms~\cite{sen2018collective}.

The works closest to ours in spirit are \citet{dietrich2005reach} and \citet{schmidtlein2022voting}. \citeauthor{dietrich2005reach} argues for maximizing acceptance treating all agents as II-proceduralists with $|R_i|=1$ based on a principle of ``Procedural Autonomy." \citeauthor{schmidtlein2022voting} demonstrates how decisions can be made by applying sequential sets of axioms, which implicitly define a rule.

The literature on constitutional amendments and ``voting on the voting rule", where agents have preferences over rules, is typically focused on types of stability or idempotency~\cite{acemoglu2012dynamics,barbera2004choosing,koray2008self} and separated into consequentialist and non-consequentialist approaches~\cite{nicolas2007dynamics}. 
 We present a unifying framework capturing both approaches because whether people accept decisions can depend on both outcomes and procedures~\cite{lind1988social,mertins2008procedural}.
 Such a framework is needed, as, for example, voters care about the fairness of electoral voting rules in tandem with their electoral preferences~\cite{plescia2020people}.

From the literature on founding and amending constitutions,~\citet{abramowitz2021beginning} put forward two constitutional mechanisms that yield an idempotent rule, along with their axiomatic characterizations.
 We demonstrate that under mild assumptions, which do not require all agents to accept a common set of axioms, both constitutional mechanisms proposed by~\citet{abramowitz2021beginning} can achieve universal acceptance.
 We do not elicit agents' preferences over menus of collective choice axioms~\cite{nurmi2015choice,de2019criterion,suzuki2020characterization} or over alternate objectives~\cite{novaro2018goal,novaro2019collective,novaro2018multi,novaro2018individual}, but instead consider preferences over rules and outcomes and the features that structure these preferences.
 Here we also mention preference models similar to our ADC model appear in \citet{bhattacharya2019constitutionally} which looks at voting over voting rules in single-peaked domains and \citep{garcia2012voting} which looks at self-selectivity of rules under trapezoidal fuzzy preferences.

 While our work focuses on preference aggregation, the basic ideas are readily extended to related areas such as judgment aggregation~\cite{list2012theory,grossi2014judgment} and binary aggregation~\cite{grandi2011binary}.

\section{Discussion and Future Work}
We have introduced a collective choice model capable of handling agent preferences over both outcomes and \emph{how} decisions are made. Our model characterizes agents by their preferences over rules and outcomes and their treatment of certain counterfactuals. Using this general model we demonstrate how to make maximally acceptable decisions and apply our approach to the concrete settings of asymmetric dichotomous choice and supermajority rule selection (amendment). Our analysis of these mechanisms focuses on the worst-case acceptance rate they achieve over a collection of instances. Remarkably, existing mechanisms achieve universal acceptance under mild conditions.

Our approach to modeling agent preferences over decisions rather than only preferences over outcomes allows us to capture a broader array of normative concerns that appear in real-world decision making but not in the existing literature. However, it is an open question how to elicit such preferences, in a fixed format or via natural language, and encode them as satisfying sets. 

An acceptance-maximizing mechanism $M$ indirectly performs approval voting over the feasible decisions using the satisfying sets as approval ballots, so agents cannot benefit by misreporting whether they accept any decision. 
However, agents may be strategic in reporting information in the profile. It is an open question how to develop mechanisms that incentivize agents to report only truthful information in both the profile and satisfying sets.

The general mechanisms we have provided are not particularly efficient for all collections of instances. Computing the outcomes of many rules on a profile can be computationally expensive. Developing more efficient application-specific mechanisms is an important open challenge.

Lastly, agents may have preferences over real-world decisions that are not reflected in our model. For example, agents might care about the computational complexity of implementing rules, whether rules are easy to understand, whether the rules preserve privacy, the probability distribution over outcomes with randomization, etc. 



\paragraph{Acknowledgements}
Nicholas Mattei was supported by NSF Awards IIS-RI-2007955, IIS-III-2107505, and IIS-RI-2134857, as well as an IBM Faculty Award and a Google Research Scholar Award. Ben Abramowitz was supported by the National Science Foundation under Grant \#2127309 to the Computing Research Association for the CIFellows Project.


\bibliography{arXiv}


\end{document}